# The development of Iranian calendar: historical and astronomical foundations

## Musa Akrami

**Department of Philosophy of Science**
**Islamic Azad University - Science and Research Branch of Tehran**
**Tehran, Iran**
E-mail: musa.akrami@srbiau.ac.ir

**Abstract.** The official Iranian calendar is a solar one that in both the length and the first day of its year is based not on convention, but on two natural (i.e. astronomical) factors: a) the moment of coincidence of the centre of the Sun and the vernal equinox during the Sun's apparent revolution around the Earth; and b) the time length between two successive apparent passages of the Sun's center across that point.

These factors give this calendar the chance that 1) its beginning is the beginning of natural solar year, 2) its length is the length of solar year, and 3) the length of its months is very close to the time of the Sun's passage across twelve signs of the Zodiac from Farvardin/Aries to Esfand/Pisces.

In this paper it would be shown that a) discussions concerning these facts have their own historical backgrounds, and b) up-to-date computations, being based on choosing the tropical year (i.e. 365.24219879 days) as the length of the calendar year, give the best possible intercalation with its specific system of leap years.

Thus, on the grounds of historical documents, astronomical data, and mathematical calculations, we establish the Iranian calendar with the highest possible accuracy, which gives it the unique exemplary place among all calendars.
**Keywords:** Solar calendrical systems, Iranian calendar, tropical year, intercalation

Introduction
1. Calendar year and the main question of the research
2. A short look at the history of Per-Islamic Iranian time reckoning
3. Calendrical systems in Iran of Islamic era
4. Jalālī calendar
5. Iranian Calendar and its fundamental elements
   1) The first year of the official calenda
   2) The length of the months
   3) The length of calendar year
   4) The problem of choosing the first day of the calendrical year
6. The system of intercalation in Iranian calendar
   1) The findings of old astronomers
   2) The researches of two modern Iranian scholars
   3) Up-to-date calculations
7. Distribution of the leap years: the best cycle, sub-cycles, and sub-sub cycles
   Method 1
   Method 2
8. The accuracy of Iranian calendar
9. Conclusion

*Introduction*

A lunar, solar, or lunisolar calendar is a device for time reckoning on the basis of two time intervals for two revolutions: the Moon's and the Sun's revolution (either real or apparent) around the Earth. Accordingly, one obtains two important time intervals, lunar month and solar year, both having several types according to the point of reference or the position of the observer. Making use of the length of the terrestrial day (i.e. the time interval of a revolution of the Earth around its axis, being 24 hours) as a measure, both time intervals are not integers, while both month and year in a calendar must contain integer numbers of days. This factor makes a proper system of intercalation necessary for a good calendar.

There are some other factors, as fundamental elements, playing important roles in calendars. We insist on four important fundamental elements in efficient calendars: 1) the length of the calendar year; this length, having a direct or indirect relation with the non-integer length of the true astronomical year, allows to determine the intercalation system of the calendar; 2) the first day of the year of the calendar; 3) the first year of the calendar; this element may be extended as the first day of the first month of the first year of the calendar (= epoch); 4) the main division of the year; a nearly 30 days month has proved to be the most appropriate division of the year according to the facts that a) the length of a solar year is nearly 12 times of a lunar month, and b) there are 12 zodiac signs traversed by the Sun in its apparent revolution around the Earth.[1] Generally, therefore, both lunar and solar years are divided into 12 months. Each month contains an integer number of days (a day being a time interval equal to 24 hours). Since the length of the year (both lunar and solar) is not an integer (i.e. an integer plus a day fraction), it does not include 12 months each with an integer number of days. But, as will be seen below, the calendar year in a sense must contain an integer number of days. This fact has led to recognize two kinds of years: ordinary year and leap year. Accordingly, the calendars need a system of intercalation to determine ordinary years and leap years.

Another important point is that the fundamental elements of the calendars may be chosen according to some conventions or/and scientific foundations. For example, in Gregorian calendar, as a calendar based on convention, the length of the year is 365.2425 days; the first day of the year is the first day of January (with no important situation relative to astronomical events); the first year of the calendar is the year in which Jesus the Christ supposedly has been born (i.e. 2,011 years before January 1, 2012); and the year is divided into 12 months (from January with 31 days to February with 28 or 29 days and September with 3o days). Thus, one may find serious difficulties with the Christian Gregorian calendar: 1) there is no relation between the first day of this calendar and the first day of natural year; 2) its system of intercalation is based on convention, having no basis in astronomical events; 3) there is no conformity between its months and natural events; 4) the length of its year is not equal to the real solar year; as we will see, this causes an error in this calendar, so that the duration needed for the error in Gregorian calendar to be added up to one day is about 3,320 years; 5) it is a national or local calendar with its particular beginning year.

In official Iranian calendar, as a calendar based on astronomical foundations, the length of the year is the length of real or mean solar year; the first day of the year is determined in accordance with the moment of coincidence of the centre of the Sun and the vernal equinox during the Sun's apparent revolution around the Earth; the first year of the calendar is the year in which the prophet of Islam has immigrated from Mecca to Medina (i.e. 622 A.D.); and the year is divided into 12 months (from Farvadin with 31 days to *Esfand* with 29 or 30 days). As we will see below, this calendar has been established on the achievements of Jalālī calendar, a calendar that has been constructed about 1079 A.D.

---

[1]. The four elements mentioned have been inferred by the author from his comparative study of result

While the three conventional fundamental elements of Gregorian calendar (i.e. the length of the year, the first day of the year, and the length of the months) make it impossible to achieve an exact and natural intercalation system, the corresponding fundamental elements of official Iranian calendar give this calendar the chance that 1) its beginning is the beginning of natural solar year, i.e. it is determined due to the moment of coincidence of the centre of the Sun and the vernal equinox; 2) the length of its year is determined in accordance with the length of solar year, i.e. the time length between two successive apparent passages of the Sun's center across that vernal equinox; and 3) the length of its months is very close to the time of the Sun's passage across the twelve signs of the Zodiac from Aries / *Farvardin* to Pisces / *Esfand* (see below).

Therefore, paying attention to both the length of different solar years defined in astronomy and other relevant astronomical events show that three important fundamental elements of the Gregorian solar calendar have been chosen according to convention, while the Iranian calendar, both in its Jalālī version and in its present official version, is a genuine solar one that in the length and the first day of its year as well as the length of its months is based not on convention, but on natural (i.e. astronomical) factors. These factors have made the Iranian calendar to be a) a natural/astronomical rather than a conventional calendar, and b) more accurate than other solar calendars such as Gregorian calendar; in fact, as many historians of science and experts in calendar studies have recognized, it has had the chance to be the most accurate calendar in the world.[2]

There is, however, another important point concerning the length of the year in the official Iranian calendar that leads to a critical point concerning its intercalation system. There have been two groups of scholars debating on the existence of a proper rule for intercalation in this calendar; 1) a group of scholars (such as Abd ul-'Ali Bīrjandī and Taghīzādeh)[3] have maintained that the length accepted for Iranian calendar is a true/real solar year, so that it is not possible to give a rule for intercalation, and the "coincidence moment" must be observed year by year and for each particular year; according to such a position, one cannot deduce calendar for future years, and having a comparative calendar for finding the equivalent dates in two or more calendrical systems is not feasible; 2) the second group of scholars, believing in the existence of some rule for intercalation, divides itself into two subgroups: 2-1) scholars who believe in a great 2,820 years cycle for intercalation that consists of its own subcycles,[4] and 2-2) scholars who believe in smaller cycles such as 161 years cycle[5] or 128 years cycle[6] or 33 years cycle.[7]

---

[2]. Nachum Dershowitz, and Edward Reingold, *Calendrical Calculations,* (Cambridge, 1997; 3rd edition, 2007), 69 and 71; Boris Rosenfeld, "Umar Al-Khayyām", *Encyclopaedia of the History of Science, Technology, and Medicine in Non-Western Cultures* (Berlin and New York, 2008)2: 479; Yunes Keramati, "Iran", *The Great Islamic Encyclopedia* (Tehran, 1368 / 1989),10: 395–396; online version: {http://en.icro.ir/index.aspx?siteid=257&pageid=8782}, accessed 2 Jan. 2012. The fact that Jalālī calendar, as the best manifestation of Iranian calendar in medieval Islamic era, has been more accurate than the Gregorian calendar has been emphasized by Western scholars many years ego (for example see H. Suter's article "Djalālī" in *Encyclopedia of Islam*, First Edition (Brill, 1913-1936),2: 1006-7; online version: Suter, H.. "Ḏjalālī" *Encyclopaedia of Islam, First Edition (1913-1936).* , 2012. Reference. 13 January 2012 <http://referenceworks.brillonline.com/entries/encyclopaedia-of-islam-1/dJalālī-SIM_2003>); Seyyed Hasan Barani, "Jalālī calendar", *Islamic Culture*, vol. 17, no. 2 (Heyderabad, 1943), 166-75; J. A. Boyle, "'Umar Khayyam: Astronomer, Mathematician and Poet", in R.N, Frye (ed), *the Cambridge History of Iran* (Cambridge, 1975), 659; Rezā 'Abdullāhī, *Tārīkh-e Tārīkh dar Iran* (= The History of Calendar in Iran) (Tehran, 1986/1365), 316.

[3]. Rezā 'Abdullāhī, *Tārīkh-e Tārīkh dar Iran* (ref.2), 316, 343; In differing with the existence of an exact rule of intercalation, Taghīzādeh cites also the decrease of the solar year length (i.e. 0.00000614 days in 100 years from 1 Jan. 1900, as shown by Newcomb); it must be said that Taghīzādeh accepts the 128 years cycle provided that one accepts the results of the present astronomical observations and computations; see Seyyed Hassan Taghīzādeh, *Gāhshomārī dar Iran-e Ghadīm* (= Calendar in Ancient Iran), (Tehran, 1938/1317) ), 172fn.

Taking into account some important revisions of Iranian calendar through its history, the author tries to explicate both mathematical and astronomical foundations of this calendar. The author has paid attention to some sources that might be classified into five types: 1) scientific documents of the past; 2) *astronomical tables* (e.g. Tūsī's *Zīj-e Īl-Khanī*, and *Zīj-e Ulugh Beg*); 3) the relevant evidence and documents remained from Jalālī Calendar; 4) the works of some contemporary scholars (e.g. Zabīh Behrūz and Ahmad Birashk); and 5) computations done by computer, using up-to-date astronomical data. It is evident that the length of the (solar) year plays a leading role in deciding on the intercalation rule. Such an emphasis leads the author to accept the tropical year (i.e. 365.24219879 days) as the length that determines the calendar year, giving an intercalation system, as the most accurate possible one, very close to that of Behrūz and Birashk[8]. Thus, the author, having his documents and grounds from the history of Persian time reckoning system and astronomical data and making use of computer, finds his intercalation system with leap years distribution very close to that of Behrūz and Birashk.[9]

*1.Calendar year and the main question of the research*

In the beginning, it might be useful to pay attention to some relevant concepts. First of all, it is an important point that the exact length of solar as well as lunar year is not an integer number. Solar year is defined as the time interval between two successive relative position of the Sun and the Earth. The astronomers have recognized several solar years according to the position of the observer and the reference point relative to which the repetition of the revolution of the Earth around the Sun occurs. Some of the important solar years are "sidereal year", "anomalistic year", "eclipse year", calendar year", and "tropical year". Among these different years, the calendar year and the tropical year have their own central role in our justified system of official time reckoning. The **tropical year** is the time interval between two successive transitions of the center of the mean Sun through a certain point in its apparent (circular) revolution around the Earth, as seen from the Earth. One may choose one of the equinoxes or solstices as the certain point suitable for measuring the tropical year.

---

[4]. Scholars such as Zabīh Behrūz and Ahmad Birashk; see Zabīh Behrūz, *Taghvīm va Tārīkh dar Iran* (= Calendar and Historiography in Iran), (Tehran, 1951/1330; 2nd Printing, 2007/1386); Zabīh Behrūz, *Taghvīm-e Nowrūzī-ye Sharyārī: Shamsī, Ghamarī* (= The Royal Nowrūzit Calendar: Lunisolar), (Tehran, 1968/1347; 2nd printing, 2007/1386); Ahmad Birashk, *A Comparative Calendar of the Iranian, Muslim Lunar, and Christian Eras for Three Thousand Years* (1260 B.H. - 2000 A.H./639 B.C. - 2621 A.D) (California, 1993).

[5].Scholars ush as Sedillot (see Seyyed Hassan Taghīzādeh, *Gāhshomārī dar Iran-e Ghadīm* (ref. 3), 169fn.) and Ali Mohammad Kaveh (see Ali Mohammad Kaveh, *Gahshomari va Tarikh-gozari Az Aghaz ta Anjam* (= Calendar and Dating from the Beginning to the End) (Tehran, 1373/1994), 110-14).

[6]. Scholars such as Reza 'Abdullāhī (see Rezā 'Abdullāhī, *Tārīkh-e Tārīkh dar Iran* (ref. 2), 345-46); as it was said in 3rd footnote, Taghīzādeh could accept this cycle on the basis of the results of the present astronomical observations and computations.

[7]. Scolars such as the authors of the *Latitude Office's Almanac of the year 1851* (see Ali Akbar Dehkhoda, "Jalālī", *Loghatnameh* (= Dehkhoda Dictionary) (Tehran, 1372), 5444) and J.F. Weidler (see Seyyed Hasan Taghīzādeh, *Gāhshomārī dar Iran-e Ghadīm* (ref. 3), 170fn).

[8]. Musa Akrami, "Computing the best intercalation for Iranian calendar", *Majalleh-ye Tārīkh-e 'Elm* (= Journal of History of Science), no. 2, (Fall 1383/2004), 88-90.

[9] . *ibid* : 81, table 4.

Since the tropical year is related to the apparent return of the Sun to the same tropic (from the point of view of an observer on the Earth), the astronomers have paid special attention to such a year from old times. "Thanks to his discovery of precession, Hipparchus, for the first time, made the distinction between the sidereal year and the tropical year".[10] He has said: "I have also composed a work on the length of the year in one book, in which I show that the solar year (by which I mean the time in which the sun goes from a solstice back to the same solstice, or from an equinox back to the same equinox) contains 365 days, plus a fraction which is less than ¼ day about 1/300th of the sum of one day and night, […]".[11]

Clearly, the tropical year maintained its central position in Ptolemy's *Almagest* too. Such an importance was transferred to the works of the astronomers of Islamic world. These astronomers tried to measure it as exact as possible. Albategnius obtained 365d + 5h + 46m + 24s, a length that was accepted by other astronomers (e.g. Mohammad Ibn Ayyūb Tabarī).[12] From middle ages up to the late nineteenth century, the European astronomers have found different values for it.

Nowadays, the value found by Newcomb has been recognized by most of the astronomers all over the world. Using the Newcomb's theory for the longitude of the Sun ($a = 129602768.13''$, $b = 1.089''$, $c = 0$), his widely known formula for tropical year is easily obtained:

$$\tau_N = 365.24219879 - 6.14 \cdot 10^{-6} \cdot T' = 365.24219265 - 6.14 \cdot 10^{-6} \cdot T \text{ days},$$

where $T' = T + 1$ and $T$ is the year elapsed from (1 January) 1900.[13] Borkowski has found a formula for the tropical year as follows:[14]

$$t = 365.242189669781 - 6.161870 \cdot 10^{-6} \cdot T - 6.44 \cdot 10^{-10} \cdot T^2$$

The author, following an old tradition of recognizing the tropical year as an appropriate accurately measurable time interval, has tried for several years to show that it is indispensable for a calendar claiming accuracy in time reckoning to accept the tropical year as the foundation of its intercalation system and determining the exact calendar years.[15] Of different values found by astronomers, the first term of Newcomb's formula is sufficiently agreeable, so that the tropical year, with a good approximation, is equal to 365.24219879 days.

The **coincidence moment**, being the exact moment of passage of the center of the (mean) Sun from a specific point in the sky,[16] may be considered as the beginning of the year in a certain calendar. The length of the **civil year** is an integer number, which is accepted as the number of the total complete days included in a year. This length may be founded on the basis of astronomical year or on convention. The **calendar year** has two meanings: i) **calendar year in its first meaning** is the exact length of the year accepted in a certain calendar that must be computed as a real or mean astronomical year; in this respect one might use two calendar year, i.e. **true calendar year** and **mean calendar year**; ii) **calendar year in its second meaning** is a year that the number of its complete days is equal to the number of the days of the civil year. This calendar year is based on the calendar year in its first meaning. Since the length of the solar year is more than 365 days and less than 366 days, a solar calendrical system must be such that one chooses 365 or 366 for the number of the days of the solar civil year and solar calendar year in its first meaning .As we will see, the number of the complete days of our solar calendar year is 365 for ordinary years and 366 for leap ones. Each calendar has a specific day as its epoch: the epoch of a calendar is the first day of the first month of the first year of that calendar.

---

[10]. J. Meeus and D. Savoie, "The History of Tropical Year", *J. Br. Astron. Assoc.* 102, 1 (1992), 40.
[11]. Ptolemy, *Almagest*, tr. by G.J. Toomer, (New Jersey, 1998), 139.
[12]. Farid Ghasemlu, *Taghvim va Taghvimnegari* (= Calendar and Calendar Compiling), (Tehran, 1388/2009), 12.
[13]. Kazimierz M. Borkowski, "The tropical year and solar calendar", *The Journal of the Royal Astronomical Society of Canada*, vol. 85, no. 3 (June 1991), 127.
[14]. *ibid*: 121.
[15]. See particularly Musa Akrami, *Gāhshomārī-ye Irānī* (= Iranian Calendar) (Tehran, 2001/1380, 2nd printing 2006/1385), 6, 78-80, 83-5, 101-2.
[16]. This moment for Iranian calendar is defined as the moment of passage of the center of the (mean) Sun through vernal equinox (i.e., traditionally, the first point of Aries).

Hence, the solar calendrical systems have two problems to be solved: 1) they must accept a particular length as the calendar year; and 2) since the number of the days of the civil year must be an integer number, the best number would be the integer number included in the solar (sidereal or anomalistic or tropical) year, i.e. 365.

The solution would be an accurate system of intercalation that would give the length of 365 days to the civil year for ordinary years, and the length of 366 days for leap years, in some big and small accurate cycles for repetition of both ordinary and leap years.

Now the main question of the research is the question that how many ordinary years would be followed by a leap year? The appropriate system of intercalation of a given calendar must solve the above problems. It is evident that if the length of the solar year were 365.25, then the length of the civil year would be 365 days for three subsequent years and 366 days for the fourth year. In such a case we could say that we would have a leap year for each K= 4 years cycle. But, the length of the solar year is not 365.25 days. So the value of K is a serious problem that is depended upon the accepted length for calendar year. Moreover, since there is no exact (i.e. moment to moment) coincidence between solar year and calendar year in its second meaning (i.e. civil year), the first day of the calendar year is not fixed.

## 2.A short look at the history of Per-Islamic Iranian time reckoning

Time reckoning and calendrical system in Iran / Persia have a rich old history, from ancient developments to modern official approval.[17] There are authoritative documents of different Iranian calendrical systems, dating from the Achaemenid period. The Old Iranian calendar was a lunisolar one, with twelve thirty days months. No direct testimony survives for the intercalation system of the Achaemenid calendar. While some scholars, e.g. Hallock, hold that the system of intercalation in o thirty days on Old Iranian calendar has been the same as that of the Babylonian calendar,[18] some others, e.g. Hartner, maintain that the intercalation system has not been the same in the Old Persian and the Babylonian calendars.[19]

---

[17].For a good report and analysis of Iranian calendar throughout its history see Seyyed Hassan Taghīzādeh, *Gāhshomārī dar Iran-e Ghadīm* (ref. 3); Seyyed Hassan Taghīzādeh, *Maghālāt-e Taghīzādeh* (=Taghīzādeh's Papers), under the supervision of Īraj Afshār (Tehran, 1970/1349),1; Antonio Panaino, "Calendars", i. Pre-Islamic Calendars, *Encyclopedia Iranica*, online version, {http://www.iranicaonline.org/articles/calendars}, accessed 16 Dec. 2011.

[18]. R. T. Hallock, *Persepolis Fortification Tablets* (Chicago, 1969), 74.

It is not difficult to find documents for caledrical parameters, e.g. the length of the solar year, in ancient Iranian texts. As an example, "different estimates for the length of the solar year in Persia may be inferred from the different statements of the *Bundahishn*" that in chapter 5 gives the length as 365 days, 5 hours, and some minutes, while in chapter 25 "contains the statement that the length of the year or 'the revolution of the sun from Aries to the end of the months' was 365 d. 6 h. and some minutes. This last estimation is also given in the *Denkard*".[20] It seems that there have been two kinds of solar year in use: a sidereal year (held to be about 365 days, 6 hours and 13 minutes) for religious purposes, and a shorter civil one for the secular affairs of the state, both requiring occasional reforms or adjustment to fix some days of solar months for the important national or religious events. Astronomical observations, yielding astronomical tables, date back to pre-Islamic era, so that "a report in the book *Az-Zīj-al-Hākimī* […] composed about the end of the tenth and the beginning of the eleventh century [centuries] by the famous astronomer Ibn Yūnis" shows that "astronomical observations were undertaken by the Persians some 360 years before the famous observations under the Abbasid Caliph al-Ma'mūn about AD 833 [i.e. about 472 and the reign of Fīrūz]".[21]

It seems that Zoroastrians have made use of a lunar year,[22] so that the Sassanid civil calendar was a lunar one with the addition of the epact in each year. Moreover, a solar calendar too was in use in which "the cumulative lag of an additional quarter-day per year was corrected, theoretically at least, by the intercalation of one month in every 120 years. [...] According to Bīrūnī [...] another system of intercalation was also used: insertion of one month in every 116 years in order to recover the quarter-days plus an additional one-fifth of an hour per year".[23] The beginning of the calendar in Achaemenid time reckoning was renewed with the first year of reigning of a new king.[24] This was an old method for specifying the first year of a calendar that the Persians had adopted from Babylonians. On the grounds of such a tradition, the year 632 A.D. [/ 10 A.P. (*Anno Persico*)[25]], i.e. the year of Yazdgerd III's sitting on the throne, was chosen by Iranian government as the beginning of a new time reckoning system, called Yazdgerdi calendar. When the Arabs conquered Iran, Zoroastrian intercalation system, manifested in Yazdgerdi calendar, with a solar year of 365.25 days, was in use. A normal year contained 365 (complete) days, and after 120 years an extra month with 30 (= 120 × 0.25) days was added. This calendar, as a Zoroastrian calendrical system with the year of reigning of Yazdgerd III (i.e. 632

---

[19]. W. Hartner, "Old Iranian Calendars", in *Cambridge History of Iran* (Cambridge, 1985), 2: 747.
[20]. Seyyed Hassan Taghīzādeh, *Old Iranian Calendars*, Royal Asiatic Society, 1938, online version, {http://www.avesta.org/taqizad.htm}, accessed 16 Dec. 2011.
[21]. Seyyed Hassan Taghīzādeh, *Old Iranian Calendars* (ref. 3).
[22]. W. Hartner, "Old Iranian Calendars", (ref. 19).
[23]. Antonio Panaino, "Calendars" (ref. 17).
[24]. Abū Reyhān Bīrūnī, *Āsār ul-Bāaghiyyeh*, translated by Akbar Danaseresht, (Tehran, 1352/1973), 48; 'Abd ul-'Alī Bīrjandī, *Sharh-e Zīj-e Jadīd-e Sultānī* (= Exposition of New Zīj of Sultān), manuscript, Libray of Iranian Islamic Consultative Assembly, no. 4716: 12b.
[25]. *Anno Persico* (= A.D.) is for the official Iranian Hijrī calendar the first year of which is 622 A.D., the year of emigration of the prophet of Islam from Mecca to Medina.

A.D.) as its beginning, played the role of the official Zoroastrian calendar in Islamic era as it is in use now among the Zoroastrians of Iran and India.[26]

Finally, it is necessary to shortly speak of the names that Iranians used for both the months and the days of a month. One may find the names of the months for Achaemenid calendar and their Old and New Elamite and Babylonian equivalents related sources.[27] What is interesting for us in this paper is the fact that the names of the months in Zoroastrian calendar have been those names[28] that have been maintained during the Islamic era, and, finally, have been officially recognized in the new Iranian (solar *Hijrī*) calendar approved in 1925 A.D. [/ 1304 A.P.] (to be described below as the official Iranian calendar).

*3. Calendrical systems in Iran of Islamic era*

Before Islam, the Arabs have been made use of a lunar calendar without paying attention to exact calculations to establish a formalized official calendrical system. It seems that the Jewish tradition of time reckoning has had its influence on the Arabian one. But the Arabian tradition, having its own names for the months, was depended upon observation of the phases of the Moon, particularly for New Moon, to accept the new month of the year. The Arabs have had some traditional ceremonies the dates of which have been determined according to their lunar time reckoning. Some of these ceremonies were religious ones with the requirement to be honored in fixed days of the year. The most important ceremony was *Hajj* that has been held performed on the tenth day of the month *Zil-Hajjah*. In fact this ceremony was at once both a religious and a national one in which the Arabs have found the chance to undertake important transactions and trades. It is easily observed that the most appropriate times for *Hajj* have been during those weeks that a) agricultural products have been gathered, and b) the weather has been moderate.

These two conditions have been realized around the fall of each solar (seasonal) year, since the variation of the seasons takes place according to solar year. But, as we know, a lunar year is about 11 days less than a solar one. This causes the 11 days anticipation of lunar months/days/ceremonies with respect to relatively fixed months/days of solar years. It is evident that obligation to respect the lunar calendar would lead to displacement of ceremonies relative to seasonal solar year, so that it was possible that *Hajj* ceremony should be performed, for example, in the heart of Winter or in the heart of Summer, with not suitable weathers and some months before or after harvest time. This has been one of the reasons for *Nasī'*, i.e. to postpone the lunar sacred months/days/ceremonies to be matched with solar fixed seasons.[29] *Nasī'* was banned by *Qur'an* at Medina, so that the lunar time reckoning was promulgate as the official sacred time reckoning method for all Muslims in all times.[30]

---

[26] Arthur Chistensen, *Iran dar Zaman-e Sasanian* (= *Sassanid Persia*), Persian tr. By Rashid Yasami, 5th printing (Tehran, 1367/1988), 532

[27] For example, see Antonio Panaino, "Calendars" (ref. 17).

[28] Hasan Ibn Ali Ghattān Marvzī, *Geyhanshenakht* (=Cosmology), facsimile copy (Qom, 1379/2000), 238.

[29] Carlo Alfonso Nallino, *Tārīkh-e Nujūm-e Islamī* (= History of Islamiv Astronomy), translated (from Arabic) Ahmad Aram (Tehran, 1349 /1970), 107-119, 121-122; Abul Fazl Musaffā, *Farhang-e Estelāhāt-e Nujūmī* (= Dictionary of Astronomical Terms), 2nd Printing (Tehran, 1366), 793.

[30] *Qur'an,* Tawbah/36-7: "Indeed, the number of months with Allah is twelve [lunar] months in the register of Allah [from] the day He created the heavens and the earth; of these, four are sacred. That is the correct religion, so do not wrong yourselves during them. […] Indeed, the postponing [of restriction within sacred months] is an increase in disbelief by which those who have disbelieved are led [further]

The *Qur'anic* emphasis on lunar months was the best ground for Muslim leaders to establish an official lunar calendar the first year of which being the year of the Prophet's immigration. The Prophet had left Mecca on Monday, 13 September 622 A.D., corresponding to 1 *Rabi' ul-Awwal* (of 1 A.H.), and 24 *Shahrivar* [(of 1 A.P.)], and had arrived to Medina on Monday, 20 September 622 A.D., corresponding to 8 *Rabi' ul-Awwal* (of 1 A.H.), and 31 *Shahrivar* (of 1 A.H.)).[31] Accordingly, under the leadership of 'Umar, the second caliph (i.e. successor of the prophet), the Arabic / Islamic time reckoning was officially recognized, seemingly after some consultation with different people including an Iranian captive commander (with the name Hormozān).[32] This calendar was established on the Arab's pre-Islamic method of time reckoning with the year of the prophet's emigration as its beginning. Afterwards the fixed beginning became prevalent in Islamic era.[33]

The Arabs succeeded to overthrow the Sassanid government in 651 A.D., and Yazdgerd III was killed; however, the Yazdgerdi calendar (with its year of 365.25 days) remained in use among the Iranian Zoroastrians. It was a seasonal calendar suitable for all seasonal ceremonies (both religious and national). Early Muslim conquerors paid no attention to calendrical systems of the conquered nations, including the Zoroastrian system. However, a lunar time reckoning inattentive to natural/solar seasons was not suitable for the vast territorial Islamic Empire with its dependence on the collection of taxes and tribute (*kharāj*, land tax in cash or commodity) mainly of agricultural products. There were different local and national solar calendrical systems in Islamic empire among various small or large circles and communities. According to about 11 days difference between solar and lunar years, using a solar time reckoning instead of lunar one proved to be necessary to fix the national and religious ceremonies, especially in tax collection affairs. An official solar calendar could allow the taxes and tribute (*kharāj*) to be collected just according to seasonal agricultural cycles. There are some reports on conversations in the second caliph's advisory council concerning the solar system of time reckoning suitable for collection of taxes in Sassanid Empire in accordance with solar calendar. It seems that the change of the beginning in Persian calendrical system has been the important factor for Persian system not to be recognized by 'Umar.[34]

Anyway, the lunar Hejrī calendar came into existence. In spite of an official acceptance of this calendar all over the Islamic empire, it seems that the situation was being prepared for making use of a kind of solar time reckoning. Biruni has a report on a dialogue between caliph Al-Motavakkel and the Zoroastrian priest concerning the time of tax collection and the necessity of intercalation to fix the festival of the first day of the year (= *Nowrūz*) and the time of paying tax and tribute[35]. The caliph Al-Mo'tazed (279-89/892-902) completed the calendrical reforms of Al-Motavakkel. Accordingly, "an intercalation of two months was introduced into the Zoroastrian year […]; through the addition of sixty days to the year 264 Yazdegerdī (282/895), *Nowrūz* [= New Day = the first day of the new year] was relocated from Saturday, 1

---

astray. They make it lawful one year and unlawful another year to correspond to the number made unlawful by Allah and [thus] make lawful what Allah has made unlawful. […]".
[31]. See Abū Reyhān Bīrūnī, *Āsār ul-Bāaghiyyeh* (ref. 24); Musa Akrami, *Gāhshomārī-ye Irānī* (ref. 15), 75.
[32]. *ibid*: 48.
[33]. Farid Ghasemlu, *Taghvim va Taghvimnegari* (ref. 12), 13.
[34]. See, for example, Abū Reyhān Bīrūnī, *Āsār ul-Bāghiyyeh*, (ref. 24), 48.
[35]. *ibid*: 51-3.

*Farvardīn* (12 *Safar* / 12 April), to Wednesday, 1 *Khordād* (13 Rabī'1/12 May)"[36] to be suitable for tax collection. Thus, in addition to Islamic lunar calendar and Yazdgerdi calendar, *Kharājī* calendar was established, with a solar year (of the length 365 days + 5 hours + 46 minutes + 24 seconds), 19 March 611 A.D. as its beginning, and Zoroastrian names for its months.[37]

*4. Jalālī calendar*

According to Kushyar Ibn Labban, the last intercalation of Yazdgerdi calendar has been done in 375 Yazdgerdi (i.e. 1006 A.D.).[38] This means that after 375 Yazdgerdi, solar Yazdgerdi year has been considered as containing 365days. Therefore, the 365.25 days year led to the retardation of *Nowrūz*.[39] As 'Abd ul-'Alī Bīrjandī has said, this retardation was equal to one day for four years (4 × 0.25 = 1), adding up to 18 days[40] in 447 Yazdgerdi[41] (1078 A.D. [/ 457 A.P.]), so that the celebration of *Nowrūz* had moved from the first of Aries / *Farvardin* to the 19[th] of Aries / *Farvardin*. Elimination of these first 18 days of *Farvardin* in the framework of a new calendar, being based on the purpose of fixing the first day of the year at the beginning of spring (i.e. vernal equinox), did seem necessary.

After some long-lasting cultural-scientific efforts for the purpose of fixing "*Nowrūz*" at the at the vernal equinox, in 467 A.H. (1074/1075 A.D.), in the reign of Jalāl ad-Dīn Malik Shāh Seljūkī (r. 465-85/1072-92), a panel of scientists (mainly mathematicians and astronomers) was entrusted the duty of revising and reforming the time reckoning in order to establish a calendar on the basis of both the length and the beginning of astronomical solar year. In his *Al-Kāmel*, Ibn Asir has spoken of 'Umar Khayyāmi, Abolmozaffar Esfazāri, and Meymūn Ibn Najīb Vāsetī as the members of the panel.[42] Some others have spoken of the participation of 'Abd ur-Rahmīn Khāzenī, Mohammad Ibn Ahmad al-Ma'rūfī al-Beyhaghī, and Abul 'Abbās Fazl Ibn Mohammad al-Lawkarī too.[43] After some years of astronomical observation and computation, a new calendar, with its own fundamental elements, was established. The resulted calendar was named in Jalāl ad-Dīn Malik Shāh's honor as "*tārīkh-e jalālī* [= Jalālī calendar], *tārīkh-e malekī*, *tārīkh-e malekšāhī*, *tārīkh-e sultānī*, and *tārīkh-e mohdas* (modern)".[44]

Accordingly, as some leading astronomers have stipulated, the principal aim of recalibrating the calendar has been to fix the first day of the year (i.e. *Nowrūz*) as the beginning of spring (the day of passing the center of the Sun from the first point of Aries.[45] Indeed, the founders of Jalālī calendar had a program with a three-partite fundamental aim: i) to find the appropriate intercalation system to fix the first day of the year; ii) to determine the best day for the first day of the year in the calendar; and iii) to make official a new calendar with its own epoch. Tūsī[46] and Ulugh Beg[47] have agreed that the first day of Jalālī calendar (i.e. Jalālī *Nowrūz* (= New Day of Jalālī calendar)) is a day in which the Sun enters the first degree of Aries until noon (i.e. before noon). On the basis of calendar reformation and in the framework of the new calendar, 18 days of retardation were eliminated as intercalation.[48] This elimination was called Sultānī / Jalālī intercalation (Kabiseh-ye Sultānī / Jalālī). Similarly, the first *Nowrūz* (i.e. the first day of the first Jalālī year, corresponding to 19th of *Farvardin* of 448 Yazdgerdi) was designated the Sultānī New Day ("*Nowrūz-e malekī*, *Nowrūz-e soltānī*, and *Nowrūz-e Hamal*"[49]), a *Nowrūz* that was to be fixed at vernal equinox due to proper computations or exact observations. The epoch (i.e. the first day of the first month of the first year) of Jalālī calendar was Friday, 1st of *Farvardin* 458 solar *Hijrī* / 9th of *Ramazān* 471 A.H.[50] (corresponding to Friday, 15 March 1079 A.D.).

---

[36]. Rezā 'Abdullāhī, "Calendars", ii. In Islamic Period, *Encyclopedia Iranica*, online version, {http://www.iranicaonline.org/articles/calendars}, accessed 16 Dec. 2011; Rezā 'Abdullāhī has cited Bīrūnī's *Āsār ul-Baghiyyeh* and Mas'ūdī's *Morūjozzahab*.
[37]. Farid Ghasemlu, *Taghvim va Taghvimnegari* (ref. 12), 58-9.
[38]. *ibid*: 62.
[39]. 'Abd ul-'Alī Bīrjandī, *Sharh-e Zīj-e Jadīd-e Sultānī* (ref. 24), 15b.
[40]. It is easy to find 18 = ((447-375) × 0.25).
[41]. 'Abd ul-'Alī Bīrjandī speaks of 448 instead of 447 Yazdgerdi (ref. 24: 16a); in fact, 448 is the beginning of the new calendar resulted from the reformation of the calendrical system.
[42]. Zabihollah Safa, *Tarikh-e Adabiyyat* (=History of Literature) (Tehran, 1336/1957),1: 311-15.
[43]. Rezā 'Abdullāhī, *Tārīkh-e Tārīkh dar Iran* (ref.2), 318, n18.
[44]. Rezā 'Abdullāhī, "Calendars" (ref. 36); 'Abdullāhī's transliteration has been changed into ours.
[45] *Nowrūz-nāme* (= the Book of the New Day), with the effort of Mojtabā Mīnuvī (Tehran, 1312/1933), 89.

The oldest report on this calendar has been written by Mohammad Ibn Ayyūb Tabarī in 476 A.H.[51] The leading references for Jalālī calendar are Tabarī's *Zīj-e Mofrad*, Khāzenī's *Zīj-e Mo'tabar-e Sanjarī*, Tusi's *Zij-e Īl-Khani*, Birjandi's *Sharh-e Zīj-e Jadid-e Sultānī*. Jalālī calendar has been included in various calendars and Persian yearbooks of the next centuries alongside other calendars. The calendar of 609 Jalālī (1687 A.D.) has been published in Rome in 1696 as the oldest sample of Jalālī calendar in Europe.[52]

Summing up, the fundamental elements of Jalālī calendar are as follows:

1) Its first year was 448 Yazgerdi / 471 A.H. [/ 458 A.P.] / 1079 A.D.;

2) Its first day of the first year was Friday, 9 *Ramazān* 471 A.H. / 19 *Farvardin* 448 Yazgerdi / 19 *Farvardin* 468 *Kharājī* / 15 March 1079 A.D. [/ 1 *Farvardin* 458 A.P.];[53]

3) Its first day of the year was determined by the exact moment of passing the Sun's center across vernal equinox; this allows to accept the solar year as the basis for calendrical year;

4) The names of its months as well as the names of the days of the months were Persian names;[54] the month names were the same as those in Yazgerdi calendar. "In order to distinguish the two calendars, in which the same Zoroastrian month names were used, the Yazdegerdī months were qualified as *qadīmī* (old) or *fārsī* and those of the Jalālī calendar as either *jalālī* or *malekī*".[55]

---

[46]. Nasīr ad-Dīn Tūsī, *Sī Fasl* (= Thirty Chapters) (Tehran, 1330 A.H. / 1912), chap. 6; Nasīr ad-Dīn Tūsī, *Zīj-e Īl-Khanī*, Manuscript, copied by Hājīshāh ibn Hossein ibn 'Abdul Munajjem Kāshānī, 890 A.H. / 1485 A.D., Iranian National Library, online copy, {http://dl.nlai.ir/UI/cbce99d9-7625-44ce-8219-c2864f3108c1/LRRView.aspx}: 36, accessed 14 Dec. 2011.

[47]. Ulugh Beg, *Zīj-e Ulugh Beg*, Manuscript, Iranian National Library, online copy, {http://dl.nlai.ir/UI/bb772138-2204-4497-9310-3308f48e6629/LRRView.aspx}: 17, accessed 15 Dec. 2011.

[48]. 'Abd ul-'Alī Bīrjandī, *Sharh-e Zīj-e Jadīd-e Sultānī* (ref. 24), 16a.

[49]. Rezā 'Abdullāhī, "Calendars" (ref. 36); 'Abdullāhī's transliteration has been changed into ours.

[50] Ulugh Beg, *Zīj-e Ulugh Beg* (ref. 47), 17. According to this *Zīj*, the epoch is not Friday 9 Ramazān 471, but Friday 10 Ramazān 471 A.H. Our findings, based both on modern tables of Ahmad Birashk and other methods of comparing calendars, show that Friday 9 Ramazān 471 is correct.

[51]. Farid Ghasemlu, *Taghvim va Taghvimnegari* (ref. 12), 60.

[52]. *ibid*: 66.

[53]. Rezā 'Abdullāhī, *Tārīkh-e Tārīkh dar Iran* (ref. 2), 298-99.

[54]. Abulfazl Naba'I, *Taghvīm va Taghvīmnegarī dar Tarīkh* (= Calendar and Calendar-making in History) (Mashhad, 1365 / 1986), 168; Rezā 'Abdullāhī, *Tārīkh-e Tārīkh dar Iran* ((ref. 2), 366-67.

[55]. Rezā 'Abdullāhī, "Calendars" (ref. 36).

5) Its ordinary years included 365 days, while its leap years included 366 days, the extra day added to 29 days of *Esfand*;

6) The calendar months were true solar month, i.e. the length of the months were equal to the time length of the passage of the Sun across the 12 signs of the Zodiac.[56]

7) According to various sources, it seems that 5 years leaps (pentaennials) have occurred for the first time in Jalālī intercalation system.

In spite of the fact that there is no exact document to directly show the Jalālī intercalation of the length of the year for Jalālī calendar, Nasīr al-Dīn Tūsī and the adherents of *Zīj-e Ulugh Beg* have spoken of  a) two kinds of leap years: i) the leap year after three ordinary years (which we call "tetraennial"), and  ii) the leap year after four ordinary years (which we call "pentaennial");[57] b) occurring one  "pentaennial" after 7 or 8 "tetraennial"s.[58] He gives the pentaennials of the first 295 years of Jalālī calendar as follows: 31, 64, 97, 130, 163, 192, 225, 258, 291;[59] the differences between two successive numbers give the cycles of occurring the pentaennials according to Tūsī's *Zīj-e Īl-Khanī*: 33, 33, 33, 33, 29, 33, and 33.

*5. Iranian Calendar and its fundamental elements*

We choose the term "Iranian calendar" (or "Persian calendar") for our special solar time reckoning method the most  important fundamental elements of which have been manifested in "Jalālī calendar" and recognized in official Iranian calendar approved by an act of the Iranian national parliament. As we told, the central points for this calendar (as probably for any other solar calendar) are the length of the year and the system of intercalation, so that the Iranian calendar has undergone different stages of acceptance and rejection with regard to its important fundamental elements during its history, particularly in Islamic era.

Iranians under the Umayyad and the Abbasid caliphs were adhering to their traditional national ceremonies of which the first day of the year (i.e. *Nowrūz*) was the most important one. While both the Umayyad and Abbasid caliphs disagree with Zoroastrian cultural heritages (including the calendar), the Abbasids had no evident opposition against *Nowrūz* and the related celebrations because of the gifts given to the rulers according to the tradition.[60] Mongols' attack led to the abolishment of the caliphs' dominance over Iran.  One of the cultural consequences of the Mongol's attack on Iran was the prevalence of two kinds of *Khānī* calendar: one with 602 A.H. (1680 A.D.) (i.e. the year in which Genghis Khān unified the Mongol tribes) as its beginning. The second kind (also called Ghāzānī calendar), with 12th of *Rajab* of 701 A.H. (corresponding to vernal equinox) as its beginning, was established according to Ghāzān Khān's command in response to the complaints of tax collectors that were in trouble by irregularities of collecting tax according to lunar  *Hijrī* calendar. According to Vābkanavī, there has been no difference between Jalālī calendar and *Ghāzānī* calendar except in the time of the beginning of the new year: while the *Nowrūz* of Jalālī calendar is the day in which the Sun enters the Aries before noon, the *Nowrūz* of the *Ghāzānī* calendar is the day in which the Sun enters the Aries till sunset.[61]

---

[56]. Farid Ghasemlu, *Taghvim va Taghvimnegari* (ref. 12), 63; As 'Abdullāhī mentions, Tūsī and Bīrjandī have confirmed this point (see 'Abdullāhī, *Tārīkh-e Tārīkh dar Iran* (ref. 2), 304-305; Rezā 'Abdullāhī, "Calendars" (ref. 36)). But 'Abdullāhī himself does not agree with such an opinion; instead, he agrees that the seasons in this calendar were astronomically true, however, as the beginning of each was marked by the apparent passage of the sun through the equinox or solstice". (see Rezā 'Abdullāhī, "Calendars" (ref. 36)).
[57]. Nasīr ad-Dīn Tūsī, *Zīj-e Īl-Khani* (ref. 46), 36; 'Abd ul-'Alī Bīrjandī, *Sharh-e Zīj-e Jadīd-e Sultānī* (ref. 24), 16a.
[58]. *Ibid.*
[59]. Nasīr ad-Dīn Tūsī, *Zīj-e Īl-Khani*, (ref. 46), 39.
[60]. 'Abdolhoseyn Zarrinkub, Tarikh-e Iran Ba'd az Islam (Histoly of Iran After Islam), vol. 1 (Tehran, 1343/1964), 453.

Islamic lunar *Hijrī* calendar, with its sacred status, maintained its prevalence particularly in religious affairs. It was recognized as the official calendar in Islamic Empire, and overshadowed other calendrical systems during the Mongols' rule. There were some efforts to officially recognize the solar (Khotanese) duodecennial animal calendar based on the central Asian animal cycle that had entered the conquered Iran. Such efforts continued even during the Safavid times. But the people resisted against this calendar, particularly having the idea that such a time reckoning had been a sign of ancient zoolatry.[62] Of course, the animal cycle survived during the centuries. One of the fundamental elements of Iranian calendar survived in these times: *Nowrūz*, i.e. the first day of the spring as the first day of calendar year, with its celebrations, a fact that had a very important role in nonofficially surviving (though weakened) of solar calendar. Though the duodecennial animal calendar with *Hijrī* starting point became officially dominant in Safavid times, the Iranians remained devoted to *Nowrūz* that was determined according to solar time reckoning. In fact, there are good clear documents from Mongols' times to *Ghājār* dynasty showing that the people have been in support of some kind of solar method of time reckoning.

After some centuries of suspense perplexity in different traditions of solar calendars and lunar (Islamic) calendar, the old difficulty with lunar year in its mismatch with seasonal year and the necessity of paying attention to the modern astronomical achievements of the West, increasingly made the solar time reckoning system important for the Iranian educated moderns. Some learned scholars such as 'Abd ul-Ghaffār Najm ud-Dowlah-ye Isfahanī and E'temād us-Saltanah were appointed by Nāser ad-din Shah to deduce a kind of comparative almanac containing different calendars. For example, 'Abd ul-Ghaffār Najm ud-Dowlah-ye Isfahanī's calendrical book contained solar *Hijrī* Calendar, lular *Hijrī* Calendar, Jewish Calendar, Sulūkī Calendar, Ghāzānī Calendar, and Yazdgeri Calendar. Around 1886 A.D. [/ 1265 A.P.], he made use of solar *Hijrī* Calendar with the Zodiac signs' names for the names of its months and the names of the animals of the duodecennial animal cycle for the names of its years repeating in a 12 years cycle.[63] Accordingly, a solar calendar based on Jalālī calendar, with the names of zodiac signs for its months, was made unofficially widespread among Iranian moderns of late Ghājār dynasty. On such a context, the triumph of Iran's Constitutional Revolution (in 1906 A.D.), as a very important modern and national event, made the improvement of the calendar necessary. Alongside the State General Audit Law, the act of solar year was ratified by Iran's National Consultative Assembly on 21 February of 1911 A.D. (i.e. the first day of Pisces (of 1289 A.P.)) as the statutory period of financial audit of the country.[64] The months of this calendar had the names of the twelve constellations of the zodiac and the years were named after the twelve animals of the duodecennial animal cycle in a cyclic order. About 14 years later, the fifth Iran's National Consultative

---

[61].Shams ud-Din Mohammad Vābkanavi, *Zij-e Mohaghghagh-e Soltani* (= Sultani Researched Zīj) Central library of Tehran University, manuscript, no. 2290, fol.53a-54b.

[62].Jean Chardin, *Siyahatname-ye Chardin* (=*Travels in Iran*), Persian text, translated by 'Abbasi (Tehran, 1338/1959), 193-97.

[63] .See 'Abd ul-Ghaffār Najm ud-Dowlah-ye Isfahanī , *Taghvīm-e Raghamī-y Sāl-e 811-ye Sijrī-ye Shamsī* (Tehran, 1267 A.P. / 1889 A.D.), {http://opac.nlai.ir/opac-prod/bibliographic/1566698}.

[64]. *Ghavanin va Ahkam* (= Laws and Sentences) (Tehran, 1318/1939), 255-56

Assembly ratified the act of the Iranian solar *Hijrī* calendar (called *Taghwīm-e Hijrī-ye Shamsī*) in April of 1925 on the basis of these fundamental elements that had some of their roots in Jalālī calendar (except in the cases of the months' names and the first year of the calendar, as seen below).[65] It seems that Seyyaed Hasan Taghīzādeh, as both an active member of the parliament and a learned researcher of the history of Iranian calendar interested in an officially recognized calendar, has played an important role in submitting the proposal of the calendar to the parliament.[66]

Paying attention to the old controversies and challenges over the existence or nonexistence of a proper rule of intercalation,[67] the author has found his own grounds to believe in the necessity of searching for a possibly exact such a rule. He, based on both the historical documents of the Iranian scholars and the astronomical achievements, has argued for the necessity of accepting mean solar year for calendar, being, in fact, the tropical year.

Now, it is the time of speaking of fundamental elements of Iranian calendar. In this calendar the number of the days of the civil year (or calendar year in its second meaning) is 365 for ordinary years; but it is 366 for leap years, occurring one time after each K or L years (we will see that K= 4, such leap years being called "tetraennial", and L = 5, such leap years being called "pentraennial").

Now we may have a look at the fundamental elements of Iranian calendar.

*1) The first year of the official calendar.* This year for present official Iranian calendar is the year of Prophet Mohammad's *Hijrah* [= emigration from Mecca to Medina]. The first day of the first year of this "solar *Hijrī* calendar" corresponds to Friday, 19 March 622 A.D.[68]

*2) The length of the months.* The length of each of its twelve months is, with very good approximation, equal to the duration of staying of the Sun in each of the twelve constellations (or signs) of the Zodiac, as has been said of Jalālī calendar by Tūsī's *Zīj-e Īl-khanī*.[69] It is interesting to compare these durations in two references: *Zīj-e Ulugh Beg* as an old document, and *Encyclopedia Britannica* as a modern reference; according to *Zīj-e Ulugh Beg*, these durations are - with my rounding the numbers for hours to zero or one, as shawn in the brackets - as follows: 1. Aries: 30 days + 15 hours [≈ 31days], 2. Taurus: 31d + 2.5h [≈ 31days] 3. Gemini: 31h + 9h [≈ 31days, or 32 days], 4. Cancer: 31d + 10h [≈ 31days, or 32 days], 5. Leo: 31d + 5h [≈ 31days], 6. Vigro: 30d +19h [≈ 31days], 7. Libra: 30d + 6h [≈ 30days], 8. Scorpius: 29d +19h [≈ 30days], 9. Sagittarius: 29d +12h [≈ 29 days or 30days], 10. Capricorn: 29d + 10h [≈ 29 or 30days] 11. Aquarius: 29d + 16h [≈ 30days], 12. Pisces: 30d + 2h [≈ 30 days].[70]

---

[65]. *Mozākerāt-e Majlis-e Showrā-ye Mellī* (= Negotiations of National Consultative Assembly),5: part 2, from Dalv 1303/ Jan.-Feb.1925 to Bahman of 1304/ Jan.-Feb. 1926 (Tehran), 1010-14, 1055-61; see also the text of the act via:
{http://www.monaseb.com/magazine-knowable/knowable-miscellaneous/306-everything-about-the-iranian-solar-calendar}, and {http://www.khabaronline.ir/news-138283.aspx}, both accessed 17 Dec. 2011.

[66] .For the discussions of the representatives see the ref. 65.

[67]. For a concise exposition of different views on the existence or nonexistence of proper rule of intercalation for Iranian calendar see Musa Akrami, *Gāhshomārī-ye Irānī* (ref. 15), 56-63.

[68]. Rezā 'Abdullāhī, "Calendars" (ref. 36). Our computations confirms not Taghīzādeh's date (i.e. 17 March 622 (Seyyed Hasan Taqīzādeh, "Various Eras and Calendars Used in Countries of Islam", *BSO(A)S* 9, (1937-39), 916)), but the date confirmed by 'Abdullāhī.

[69]. Rezā 'Abdullāhī, *Tārīkh-e Tārīkh dar Iran* (ref.2), 305.

[70]. Seyyed Jalāl ad-Din Tehrānī, *Gāhnāme-ye 1313* (= the Calendar of the Year 1313) (Tehran, 1934/1313), 54-5.

According to recent findings,[71] vernal equinox is circa 21$^{st}$ or 22$^{nd}$ of March (with 80 days from the beginning of the year), while autumnal equinox is circa 23$^{rd}$ or 24 of September (with 266 days from the beginning of the year. Thus, the length of the first six months of solar year is 266 − 80 = 186 days, which is equal to the sum of the lengths of the six months of Iranian solar calendar from the first day of *Farvardin* / Aries to the last day of *Shahrivar* / Vigro. A looking at the length of the months as rounded by me in the brackets, we would be justified by anonymous *Rabi' Al-Monajjemīn* that accepts 31 days for 1$^{st}$, 2$^{nd}$, 4$^{th}$, 5$^{th}$, and 6$^{th}$ months, 32 days for 3$^{rd}$ month, 30 days for 7$^{th}$, 8$^{th}$, 11$^{th}$, and 12$^{th}$ months, and 29 days for 9$^{th}$ and 10$^{th}$ months.[72]

*3) The length of calendar year.* The length of the year for Iranian calendar is equal to solar year (i.e. the time length of the revolution of the Earth around the Sun). It may be possible to accept one of astronomical solar years. Both the kind of the accepted solar year and its exact length have been controversial during the history of calendar studies.

*4) The problem of choosing the first day of the calendrical year.* The first day of the calendrical year is the same as the first day in which the passage of the Sun's center through the vernal equinox takes place.

The act of Iranian calendar, ratified in 1925, has approved above four fundamental elements. There is no vagueness in relation to the 1$^{st}$ and 4$^{th}$ fundamental elements. These two fundamental elements have been accepted without any doubt and ambiguity. The months of this calendar, with their Persian names of the Zoroastrian calendar that had been maintained during the Islamic era (as cited above from Ghattān Marvzī's *Geyhānshenākht*), are as follows:

1. *Farvardin* (corresponding to Aries, March - April):31 days
2. *Ordibehesht* (corresponding to Taurus, April -May): 31 days
3. *Khordad* (corresponding to Gemini, May-June): 31 days
4. *Tir* (corresponding to Cancer, June-July): 31 days
5. *Amordad* (corresponding to Leo, July-August): 31 days
6. *Shahrivar* (corresponding to Vigro, August-September): 31 days
7. *Mehr* (corresponding to Libra, September-October): 30 days
8. *Āban* (corresponding to Scorpius, October-November): 30 days
9. *Āzar* (corresponding to Sagittarius, November-December): 30 days
10. *Dey* (corresponding to Capricorn, December-January): 30 days
11. *Bahman* (corresponding to Aquarius, January-February): 30 days
12. *Esfand* (corresponding to Pisces, February-March): 29 days in ordinary years, 30 days in leap years.[73]

An important point in the act of the Iran's parliament is that the calendar year is a "**true solar year**". The act is silent on the question of the intercalation system, a matter of fact that showed its importance very soon. The problem of the appropriate system of intercalation, however, remained unsolved in spite of some efforts to postulate such a system. It is evident that "true solar year" is not fixed. Therefore, it is not suitable for calendar year: calendar year must be fixed to make comparison

---

[71]. United States Naval Observatory (2010-06-10), "Earth's Seasons: Equinoxes, Solstices, Perihelion, and Aphelion, 2000-2020"; {http://www.usno.navy.mil/USNO/astronomical-applications/data-services/earth-seasons.}, accessed 16 Dec. 2011.

[72]. Amir Fereydūn Garakānī, *Gāhnāme, Rūz-e Yekom-e Farvardin, Nowrūz* (= Calendar, the first day of Farvardin, New Day) (Tehran, 1976/1355), 30.

[73]. See the text of the act of the solar calendar via {http://www.khabaronline.ir/news-138283.aspx}

between the dates of different calendars as well as determining the future dates possible.

## 6.The system of intercalation in Iranian calendar

As has been mentioned, the author has made use of the findings of old astronomers, the researches of two modern Iranian scholars, and up-to-date calculations to establish and prove the proper system of intercalation for Iranian calendar.

*1) The findings of old astronomers.* I showed that, according to Tūsī and the adherents of *Zīj-e Ulugh Beg*, we are allowed to accept the existence of a) both "tetraennial"s (i.e. four years leaps) and "pentaennial"s (i.e. five years leaps), and b) 29 years and 33 years cycles of occurring the pentaennials in Jalālī calendar.

*2) The researches of two modern Iranian scholars.* These two scholars are Zabīh Behrūz[74] and Ahmad Birashk.[75] They have accepted a system of intercalation based on a 2820 years principal cycle with its own 128 years sub-cycles, 29 or 33 years sub-sub-cycles, "tetraennial"s, and "pentaennial"s.

*3) Up-to-date calculations.* In our calculations, we have accepted the length of tropical year as the most appropriate length for calendar year. As we have told, the first term of Newcomb's formula, i.e. 365.24219879 days (for the first day of January 1900), has been accepted as tropical year and mean solar year.

No, choosing the tropical year as both the mean solar year and the calendar year in its first meaning, we must find a system of intercalation that the average of civil years during one principal cycle becomes equal to (or, with the best approximation, very close to) tropical year. As was said, tropical year is equal to 365.24219879 days (≈365 days + 5 hours + 48 minutes + 45.975456 seconds).

It may be written as "tropical year" = 365 days + "day fraction", in which the "day fraction" is equal to 0.24219879 days (= 5 hours + 48 minutes + 45.975456 seconds).

We must note that if the coincidence moment, being defined as the moment of passage of the (mean) Sun's center through vernal equinox, is before the midday (i.e. noon) of a certain day, then the same day would be the first day of the "new year", and the elapsed year is an ordinary year.

Since i) the difference between two subsequent coincidence moments is equal to tropical year, and ii) the midday is at 12 o'clock, we, subtracting the "day fraction) from 12 hours, may find the condition for leap/ordinary years. First of all, we find the difference between midday and the "day fraction":

$A = 12$ hours − "day fraction" ≈ 6 hours + 11 minutes + 14.024544 seconds = 6:11:14.024544.

Suppose that the beginning [= coincidence moment] of a given year is B. Then,
i) the given year is a leap year if B is inside the interval (A,B) i.e. $A < B < 12$.
ii) the given year is an ordinary year if B is outside the interval (A,12], i.e. $B \leq A$ and $B \geq 12$, from 12:00:00 to 6:11:14.024544.

---

[74]. Zabīh Behrūz, *Taghvīm va Tārīkh dar Iran* (ref. 4); Zabīh Behrūz, *Taghvīm-e Nowrūzī-ye Sharyārī: Shamsī, Ghamarī* (ref. 4).
[75]. Ahmad Birashk, *A Comparative Calendar of the Iranian, Muslim Lunar, and Christian Eras for Three Thousand Years* (ref. 4).

*7.Distribution of the leap years: the best cycle, sub-cycles, and sub-sub cycles*

Now, we must find a cycle during which the coincidence moment would be repeated as exactly as possible. We will see that the most appropriate cycle is **2,820 years** one. This cycle could be found via two methods.

*Method 1.* In this method we insist on some bases: 1) if the number denoting the cycle is multiplied by tropical year, the result would be as close to an integer number as possible; 2) since the coincidence moment of the years of the beginning and the end of the cycle are at the midday, and the coincidence moment of the midcycle year is at the midnight, the cycle must be an even number; 3) we must choose a number that its sub-cycles and sub-sub-cycles have past records in the history of science. Calculations show that the best number is 2820: 365.24219879 × 2820 = 1029983.0005878 days. 0.0005878 days is about 51 seconds (a very small number, that we may neglect it as a good approximate).

*Method 2.* Using tropical year, it is possible to write a computer program to form an arithmetic progression, with 12 as the first term and the day fraction as the common difference. This procedure would yield the most accurate calendar with the best cycle which begins from the beginning of the year that its coincidence moment is at the midday (i.e.12 o'clock). Using the definition of coincidence moment and leap year, the author has managed to calculate the leap years by computer. The cycle as well as its sub-cycles and sub-sub-cycles would be found during the process of the calculation.

The coincidence moment of the 1$^{st}$ year of the cycle = 12 + **0**×0.24219879 = 12:00:00

The coincidence moment of the 2nd year of the cycle = 12 + **1**×0.24219879 = 17:48:45.975456

The coincidence moment of the 3rd year of the cycle = 12 + **2**×0.24219879 = 23:37:31.950912

The coincidence moment of the 4th year of the cycle = 12 + **3**×0.24219879 = 05:26:17.926368

The coincidence moment of the 5th year of the cycle = 12 + **4**×0.24219879 = 11:15:3.901824

The coincidence moment of the 6th year of the cycle = 12 + **5**×0.24219879 = 17:03:49.87728

According to our criterion, as mentioned above, the 5th year is a leap year, while the other years are ordinary ones. So the cycle begins with a pentaennial.

The calculation will show that this first pentaennial would be followed by 6 tetraennials. That is, starting from the beginning of the cycle (with a starting point at the noon of the first day of the first year), the following years would be leap years: 5th, 9th, 13th, 17th, 21st, 25th, and 29th. These leap years would make a 1×**5** + 6×**4** = **29** years sub-sub-cycle. Afterwards, we would see another pentaennial and 7 tetraennials. Accordingly, the next sub-sub-cycle would be a **33** (= 1×**5** +7×**4)** years one. That is, in a 33 years sub-sub-cycle, the following years would be leap years: 5th, 9th, 13th, 17th, 21st, 25th, 29th, and 33rd.

The first 29 years sub-sub-cycle would be followed by three 33 years sub-sub-cycles. After these four sub-sub-cycles (i.e. one 29 years sub-sub-cycle and three 33 years sub-sub-cycles) we will observe another 29 years sub-sub-cycle followed by three 33 years sub-sub-cycles. Accordingly, it seems that there is a **128** (=1×29 + 3×33) years sub-cycle. Yes! This is so! But sometimes we will observe that the number of 33 years sub-sub-cycles would be not 3 but 4, so that there would be a **161** (=1×29 + 4×33) years sub-cycle. Computation will show that the 2,820 years cycle will be completed with following sub-cycles: 1×**128** years + 4 (4×**128** years + **161** years) = **2,820** years.

Now we summarize the cycle as follows:

1×**5** = **5** (the first pentaennial occurring at the beginning of a 29 or 33 years sub-sub-cycle)

6×**4** = 24 or 7×**4** = 28 (tetraennials occurring in a 29 or a 33 years sub-sub-cycle)

1×**5** + 6×**4** = **29** (the first sub-sub-cycle occurring in a 128 years sub-cycle)

1×**5** + 7×**4** = **33** (33 years sub-sub-cycles occurring in a 128 years sub-cycle)

1×**29** + 3×**33** = **128** (a kind of sub-cycles occurring in the 2820 years cycle)

1×**29** + 4×**33** = **161** (another kind of sub-cycles occurring in the 2820 years cycle)

1×**128** + 4×**128** + 1×**161** + 4×**128** + 1×**161** + 4×**128** + 1×**161** + 4×**128** + 1×**161** = **128** + 4 (4×**128** + **161**) = **2,820** (the complete cycle).

We find the 2,820 years cycle in Behrūz, and, following him, Birashk. Behrūz gives some historical evidence and documents for his finding. We have no idea concerning his documents and arguments. Of course, there is a difference between the result of our calculation and Behrūz / Birashk achievement: there is no 161 years sub-cycle in their 2,820 years principal cycle, leading some differences in the distribution of leap years. Their 2,820 years cycle in terms of sub-cycles and sub-sub-cycles is as follows:

**2,820** = 22×**128** +1× **4** = 22× (1×**29** + 3×**33**) +1× **4** = 22× [(1×**5** + 6×**4**) + 3(1×**5** + 7×**4**)] +1× **4**

The difference in the distribution of leap years leads to 16 differences in the places of tetraennials and pentaennials, that may be neglected as a good approximation.

Now one might criticize Tūsī and the adherents of *Zīj-e Ulugh Beg* in respect to two points: i) as it is evident, our result is in opposition to the opinion of Tūsī and the adherents of *Zīj-e Ulugh Beg* about the place of pentaennials, i.e. not in the beginning of the 29 years and 33 years sub-sub-cycles, but in the end of them; ii) they have spoken of occurring one "pentaennial" after 7 or 8 "tetraennial"s[76] (i.e. 7×**4** + 1×**5** = **33** years and 8×**4** + 1×**5** = **37** years), while, as we told, the cycles of occurring the pentaennials according to Tūsī's *Zīj-e Īl-khanī* are 33, 33, 33, 33, 29, 33, and 33 in which there is no 37 years cycle; instead there is a 29 years cycle in which there is one "pentaennial" (of course not after but before 6 "tetraennial"s) without being expressed explicitly.

*8. The accuracy of Iranian calendar*

We need the number of leap years of a 2,820 years cycle in order to calculate the day fraction for Iranian calendar. In spite of 16 differences in the places of tetraennials and pentaennials, there is no difference in the number of both leap and ordinary years among our computation and Behrūz / Birashk achievement.

The distribution of the leap years in our 2,820 years cycle is as follows:

**2,820** = [(1×**5** + 6×**4**) + 3(1×**5** + 7×**4**)] + 4{4[(1×**5** + 6×**4**) + 3 (1×**5** + 7×**4**)] + [(1×**5** + 6×**4**) + 4(1×**5** + 7×**4**)]}

Accordingly, it is possible to find the number of leap/ordinary years of the cycle:

The number of leap years of a 29 years sub-sub-cycle = 1 + 6 = **7**;

The number of leap years of a 33 years sub-sub-cycle = 1 + 7 = **8**

---

[76]. Nasīr ad-Dīn Tūsī, *Zīj-e Īl-Khani* (ref. 44), 36; 'Abd ul-'Alī Bīrjandī, *Sharh-e Zīj-e Jadīd-e Sultānī* (ref. 24), 16a.

The number of leap years of a 128 years sub-cycle = 1×**7** +3×**8** = **31**
The number of leap years of a 161 years sub-cycle = 1×**7** +4×**8** = **39**
The number of leap years of the 2820 years cycle = **31** + 4(4×**31** + **39**) = **683**
The number of ordinary years of the 2,820 years cycle = 2,820 – 683 = **2,137**

These numbers, for leap and ordinary years, are seen in Behrūz / Birashk 2,820 years cycle too. Now we may find the day fraction and mean calendar year: DF = 683 ÷ 2,820 = 0.24219858

Mean calendar year is equal to 365 + 0.24219858 (very close to tropical year, according to both Newcomb and Borkowski). Of course, such a result is not a surprising for our method in which we have started with day fraction of Newcomb's tropical year (i.e. the first term of the Newcomb's formula). But one may find it interesting that Behrūz has not directly made use of a day fraction equal to or close to day fraction of tropical year: he, apparently, has had found some historical evidence for his cycle, sub-cycle, and sub-sub-cycles.

One minute (i.e. 1 ÷ (24×60) = 0.00069444 days) of error would add up to one day (i.e. 24 hours) in 1,440 (= 1 ÷ 0.00069444 = 1 ÷ (24×60)) years. Since the Christian Gregorian Calendar year is 365.2425 days,[77] the accuracy of Christian Gregorian Calendar may be found according to its DF which is 0.2425: 1day ÷ (365.2425days – 365.24219879 days) / year ≈ **3,320** years (in spite of the fact that this calendar has been come out from some corrections in Julian calendar with a year of 365.25 days, implying it to be corrected for one day in 128 years[78]).

Accepting the 2,820 years cycle, with one of the above distributions for leap years, one finds the difference between our mean calendar year and tropical year: 365.24219879days – 365.24219858days = 0.00000021day. The error will add up to 0.0005878 days or 51seconds in each 2,820 years cycle.[79] This possible error will add up to one day in about **4,761,905** years: 1day ÷ 0.00000021day / year ≈ **4,761,905** years.

Accordingly, even if the error for Iranian calendar is to be considered as significant, this calendar is more accurate than Christian Gregorian Calendar up to 1,434 times: 4,761,905 ÷ 3,320 ≈ **1,434**

Now it is the turn to repeat the confession of some Western scholars concerning the accuracy of Iranian calendar with respect to other calendars including the Christian Calendar. Recently, N. Dershowitz and E.M. Reingold have argued that "the modern Persian Calendar […] is an extremely accurate solar calendar"[80] with "amazing accuracy".[81]

## 9. Conclusion

Calendars are defined as time reckoning systems. For us, as the residents of the Earth, they usually are based on either the revolution of the Moon around the Earth or the Sun's (apparent) motion around the Earth. Their time reckoning systems generally are based on a year length, a first day of the year, and a beginning year. It is evident that there are serious difficulties both in relation to the errors occurred in most existent calendars and in comparing the dates in two calendars.

---

[77]. Peter Meyer, "The Julian and Gregorian Calendars", online version, {http://www.hermetic.ch/cal_stud/cal_art.html}, Retrieved 24 Sep. 2011.
[78]. Musa Akrami, *Gāhshomārī-ye Irānī* (ref. 15), 35-6, 41.
[79]. Obviously, the error is less than 0.0005878 if one makes use of Brokowsky's formula for tropical year.
[80]. Nachum Dershowitz, and Edward Reingold, *Calendrical Calculations* (ref. 2), 69.
[81]. *ibid*: 71.

According to the mismatch of lunar calendars with seasonal year, they cause some difficulties in comparison with solar calendars. Thus the calendars of interest in this paper are the solar ones. These calendars are of two kinds: 1) those established according to astronomical proper events, and 2) those constructed on some conventions or contracts among the people of a society. The conventional character of the latter kind, without any direct coincidence with astronomical year and natural events, has been a negative point that Iranians have tried to abandon.

Comparative study of different calendars of the past and the present shows a spectrum of accuracy according to the fact that how close their years are to real solar year. Another important fact in time reckoning is the moment of passage of the center of the (mean) Sun from the vernal equinox.

Since the solar year length is not an integer number (i.e. not having integer number of days) but has a day fraction, it has been needed to apply an exact or approximate intercalation method. In this regard, the scholars, from ancient times up to the present, have been divided into two groups: i) those who have not accepted any proper rule for computing the kind of the future years beforehand, and ii) those who have accepted a proper rule. Of course, these scholars themselves have suggested different rules.

On the grounds of both historical documents and astronomical data, the author believes in the existence of a proper rule for intercalation, and argues for the necessity of accepting a) the tropical year as the "mean solar year" suitable for finding an accurate intercalation method and a good calendrical system, on the one hand, and b) the moment of coincidence of the center of the Sun and vernal equinox as the proper moment for distinguishing the new (coming) year from the old (passing) one.

The author has made use of some historical documents (with critically analyzing them), the achievements of modern researchers of the field of calendar studies, modern astronomical findings, and computer calculation to argue for the fundamental importance of three fundamental elements of the Iranian calendar as the most accurate model for the solar calendars: the best calendar is one that 1) its first day of the year is the same as the first day of astronomical year (e. i. the day specified on the basis of the coincidence of Sun's center and vernal equinox); It has been shown in the paper that the best first day of a calendar is the day in which the passage of the Sun's center from the vernal equinox occurs between 6:11:14.024544 and 12:00:00 (midnight being 00:00:00); 2) its year length is determined according to tropical year as the mean solar year; 3) the main division of its year, i.e. division into months, is in accordance to the time interval needed for the Sun to transverse across each corresponding Zodiac sign. The first two fundamental elements guarantee to establish the best intercalation system and the most accurate possible calendar. The conformity of the calendar with natural (seasonal) year becomes complete by the third fundamental element.